# Atomic resolution mapping of localized phonon modes at grain boundaries

Benedikt Haas[1*], Tara M. Boland[2†], Christian Elsässer[3], Arunima K. Singh[4], Katia March[5], Juri Barthel[6], Christoph T. Koch[1] and Peter Rez[4]

[1] Department of Physics & IRIS Adlershof, Humboldt-Universität zu Berlin, 12489 Berlin, Germany
[2] School for Engineering of Matter Transport and Energy, Arizona State University, Tempe, AZ 85287-6106, USA
[3] Fraunhofer Institute for Mechanics of Materials IWM, 79108 Freiburg, Germany
[4] Department of Physics, Arizona State University, Tempe, AZ 85287-1504, USA
[5] Sorbonne Université, Muséum National d'Histoire Naturelle, UMR CNRS 7590, Institut de Minéralogie, de Physique des Matériaux et de Cosmochimie, IMPMC, 75005 Paris, France
[6] Ernst Ruska-Centre (ER-C 2), Forschungszentrum Jülich, 52428 Jülich, Germany
*Corresponding author. Email: haas@physik.hu-berlin.de
†Present address: Computational Atomic-Scale Materials Design (CAMD), Technical University of Denmark; Kgs. Lyngby, 2800, Denmark

ABSTRACT

Phonon scattering at grain boundaries (GBs) is significant in controlling nanoscale device thermal conductivity. However, GBs could also act as waveguides for selected modes. To measure localized GB phonon modes, meV energy resolution is needed with sub-nm spatial resolution. Using monochromated electron energy loss spectroscopy (EELS) in the scanning transmission electron microscope (STEM) we have mapped the 60 meV optic mode across GBs in silicon at atomic resolution and compared it to calculated phonon densities of states (DOS). The intensity is strongly reduced at GBs characterised by the presence of five- and seven-fold rings where bond angles differ from the bulk. The excellent agreement between theory and experiment strongly supports the existence of localized phonon modes and thus of GBs acting as waveguides.

MAIN

Grain boundaries (GBs), the interfaces between crystallites that make up polycrystalline microstructures, control many materials properties. The scattering of phonons at GBs, especially localized modes, is significant for thermal conductivity both at the nanoscale and for polycrystals at the macroscale. In particular thermal conductivity at interfaces has been a significant problem for heat dissipation in nanoscale devices [1]. It is also critical for improving the performance of thermoelectrics that allow for the direct conversion of heat to electrical energy with no moving parts. The thermal conductivity is directly proportional to the phonon relaxation time that is related to phonon scattering from defects. GBs predominately scatter low frequency phonons that control thermal conductivity over critically important temperature ranges [2,3].

Recently, there has also been interest in using phonons rather than photons for coherent coupling between mechanical oscillators as a new paradigm for quantum sensing and information processing. Zivari et al. demonstrated phonon propagation in a single mode cavity waveguide fabricated in silicon [4]. Grain boundaries could act as waveguides that select particular localized phonon modes. These guided phonons could propagate further than in the bulk, thus opening up the possibility of building phononic devices. To detect and characterize such modes requires atomic resolution for the selected mode.

So far investigations of GB effects on phonons have either been purely computational or used techniques such as Raman spectroscopy that lack the spatial resolution to detect individual vibrational modes at the GB. Developments in monochromated electron energy loss spectroscopy (EELS) in the scanning transmission electron microscope (STEM) have made it possible to probe phonon modes at the nanometer scale [5], and even at atomic resolution in some well-chosen cases [6,7]. Hage et al. measured vibrational modes from a single silicon atom substituting for carbon in a graphene lattice [8]. Yan et al. showed the changes to vibrational spectra from a stacking fault in SiC [9] and Gadre et al. showed how the optic mode changes for sharp and diffuse interfaces between Ge and Si in a GeSi quantum dot, and related their observations to phonon dynamics [10]. Monochromated EELS in the STEM has also been used to probe how phonon modes change at heterointerfaces between diamond and hexagonal BN (hBN) [11], and Si and AlN [12]. Recently Hoglund et al. reported changes in the optic mode spectra at different positions in a 10° boundary in strontium titanate, though not at atomic resolution [13].

Here, we demonstrate atomic resolution phonon mapping of buried defects, in our case a symmetric Σ3 (111), a symmetric Σ9 (221) and an asymmetric (111)|(115) GBs at a junction of three grains in a silicon polycrystal. The experimental spectra are compared to simulated EELS spectra obtained from structure optimization of atomic models using molecular dynamics (MD) simulations, and subsequent computation of the EELS spectra directly from the MD trajectories. For the Σ3 (111) GB a model was built from high-resolution high-angle annular dark-field (HAADF) data acquired during the experiments while for the symmetric Σ9 (221) and asymmetric (111)|(115) GBs pre-existing models from literature are used (from Stoffers et al. [14] and Ziebarth et al. [15], respectively) that were checked for consistency with the HAADF images.

Fig. 1a is a low-magnification bright-field STEM image of a focused ion-beam prepared silicon sample in [-110] orientation exhibiting the Σ3, Σ9 and (111)|(115) boundaries. Dark areas originate from carbon contamination that arose during previous experiments in a non-UHV microscope [14] and have been avoided in the present study. HAADF images acquired for the symmetric Σ3 (111), the symmetric Σ9 (221) and the asymmetric (111)|(115) GBs are depicted as Fig. 1b-d, respectively, from the areas marked with stars in Fig. 1a. A simplified diagram of the EELS experiments performed in a Nion HERMES STEM with IRIS spectrometer is shown in Fig. 1e. A convergent electron probe is focused on the sample and electrons scattered over a range of angles defined by the spectrometer entrance aperture are collected. These electrons are dispersed in energy by the prism of the spectrometer and recorded using a direct electron detector. In our experiments the primary beam energy was 60 keV, the probe on the specimen had a 30 mrad convergence semi-angle and the spectrometer collection semi-angle was 42 mrad. Further experimental parameters are given in the Supplementary Materials. A spectrum from bulk Si along [-110] is shown in Fig. 1f with the blue graph depicting the raw data and the orange one the result after subtracting the zero-loss peak (ZLP), revealing the phonon signatures.

Since Si has two atoms in each primitive unit cell it has both an optic and acoustic mode. However, there is no oscillating dipole that results in a spatially delocalized signal because the two atoms are identical. Following Venkatraman et al. [7] and Gadre et al. [10] it is then possible to collect the EELS signal on axis which results in a much stronger signal than would be collected with the displaced collection aperture required when dipole scattering is significant to not obscure atomic resolution through delocalization [6].

The silicon optic mode with a relatively flat dispersion gives a peak in the density of states at about 60 meV and flat regions of the acoustic mode near the Brillouin Zone boundaries result in a smaller peak at 40 meV.

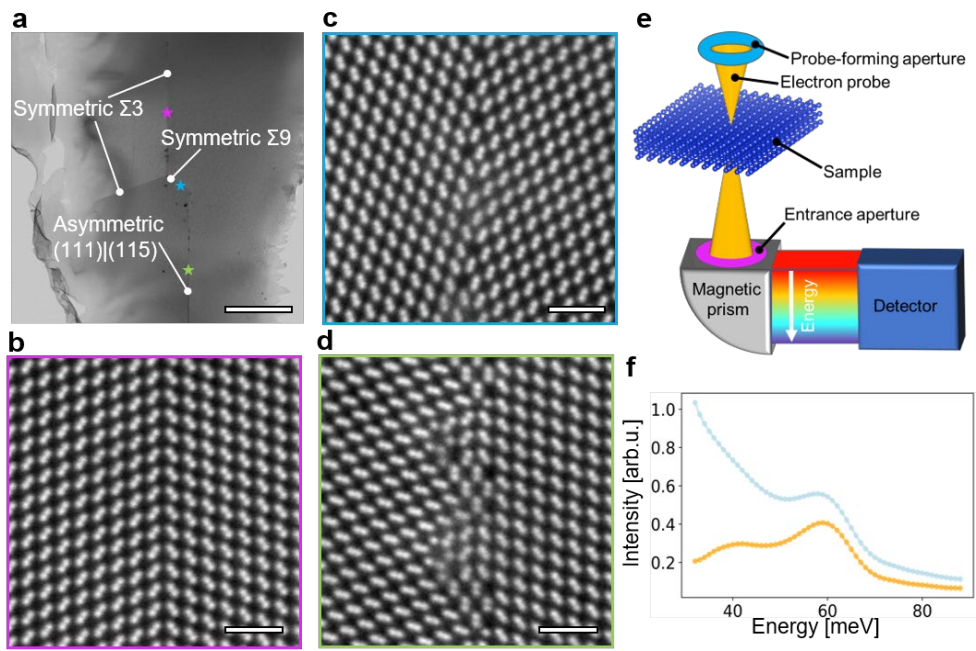

**Fig. 1 | Structure of GBs and spectroscopy setup. a**, Overview bright-field image of the silicon sample showing the different GBs. Dark spots in the overview map stem from carbon contamination from previous experiments in a non-UHV microscope [14]. Scale bar, 500 nm. **b-d**, High-resolution HAADF images of the Σ3 (111), symmetric Σ9 (221), and asymmetric (111)|(115) GBs, respectively. The positions of acquisitions are marked in **a** as stars. Scale bars, 1 nm. **e**, Experimental setup for the EELS measurements. **f**, Experimental spectra from bulk silicon before (blue) and after (orange) background-subtraction, revealing the phonon signals at 40 meV and 60 meV.

Calculated phonon dispersion and densities of states (DOS) of phonons in crystals can be obtained by diagonalization of the dynamical matrix using force constants computed by methods of density functional theory (e.g. with the codes VASP [16] and Phonopy [17]) and the results agree well with experimental measurements [17-19]. However, this approach is computer resource intensive and is not practical for systems with large numbers of atoms such as the supercells representing GB structures. Instead we take the Fourier transform of the velocity-velocity correlation function of MD trajectories computed using the Tersoff [20] empirical potential. It has been shown that this gives equivalent results, though the peaks are shifted to slightly higher energies with this empirical potential [21].

Fig. 2a-c show HAADF images that were acquired in parallel with EELS maps of the Σ3 (111), symmetric Σ9 (221), and asymmetric (111)|(115) GBs, respectively. The change in appearance compared to the high-resolution HAADF images of Fig. 1b-d stem from different experimental parameters (see Methods and Fig. S1). In Fig. 2d-f, the amplitude of the 60 meV peak is mapped out for the three GBs. In Fig. 2g-i spectra from the regions indicated by squares in Fig. 2d-f for the GB and bulk on both sides are shown. While almost no variation is visible for the case of the Σ3 (111) GB, a clear dip in intensity is apparent for the symmetric Σ9 (221) GB which is even more pronounced for the asymmetric (111)|(115) GB. Variation of the phonon DOS is even clearly correlated with the structure along the boundary for the asymmetric (111)|(115) GB in the map. This correlation exists also for the Σ9 (221) GB, which can be better seen when extracting the averaged structural motif via template-matching (cf. Fig. S2).

To ensure that indeed variations of the phonon DOS are mapped and not geometric effects due to being on or off an atomic column, the intensity of the peak around 40 meV was also investigated (see Fig. S3). Since geometric effects would influence the relative intensity variations of the 40 meV peak in a similar manner, they can be ruled out.

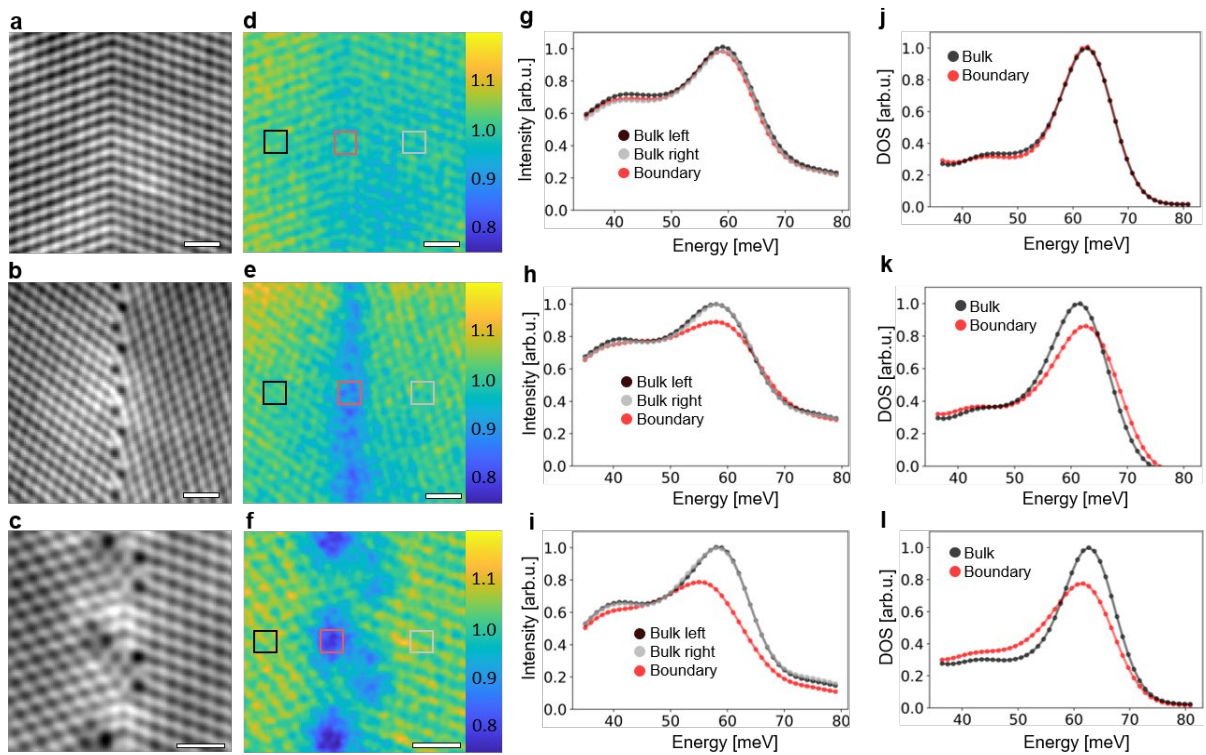

**Fig. 2 | Comparison of experimental and calculated spectra for different GBs. a-c**, HAADF images acquired in parallel with spectrum maps of Σ3 (111), symmetric Σ9 (221), and asymmetric (111)|(115) GBs, respectively. Scale bars, 1 nm. **d-f**, Maps of the amplitude of the 60 meV phonon peak, corresponding to the images in **a-c**, respectively. Scale bars, 1 nm. **g-i**, Spectra of bulk regions and from the GB for the three different GBs. **j-l**, Calculated spectra for the three types of GBs.

Calculated phonon DOS that are derived from MD of structural models, based on or confirmed by the HAADF images in Fig. 1b-d as mentioned above, are given in Fig. 2j-k. Good agreement between calculated and experimental spectra can be observed for the peak around 60 meV. Table 1 shows quantitative values for the height ratio and also the energy shift of the 60 meV peak for GB atoms relative to the same peak from bulk regions, shown as red and black boxes respectively in Fig. 2d-f. The uncertainties were determined by measuring peak height and position over similar 10 by 10 pixel regions at different positions in the bulk (both sides of the GB were sampled) and calculating the standard deviation. The calculated depression of the 60 meV peak at the different boundaries is in excellent quantitative agreement with experimental results.

**Table 1 | Experimental and calculated height and position of 60 meV peak for different GBs in Si.**

| GB Type | Experimental GB to bulk ratio | Calculated GB to bulk ratio | Exp. GB to bulk shift (meV) | Calc. GB to bulk shift (meV) |
|---|---|---|---|---|
| Σ3 (111) | 0.99 ± 0.03 | 1.01 | (-0.3 ± 0.2) | -0.1 |
| Σ9 (221) | 0.89 ± 0.02 | 0.86 | (0.0 ± 0.3) | +1.1 |
| Asymmetric (111)\|(115) | 0.79 ± 0.01 | 0.78 | (-3.2 ± 0.2) | -1.3 |

The detailed theory for the intensity of features related to phonon scattering in STEM EELS is given by Rez and Singh [22] and Zeiger and Rusz [23]. It would seem that the dynamical diffraction of the probing and scattered electrons would have the same effect for spectra acquired from boundary and bulk regions. For the position of the 60meV peak, there is good agreement for Σ3 (111) and reasonable agreement for the asymmetric (111)|(115) GB while for the symmetric Σ9 (221) GB a +1.1 meV shift is expected but none is observed. As discussed by Rez et al. [21] any shifts could depend on the nature of the empirical potential so perfect agreement is not expected.

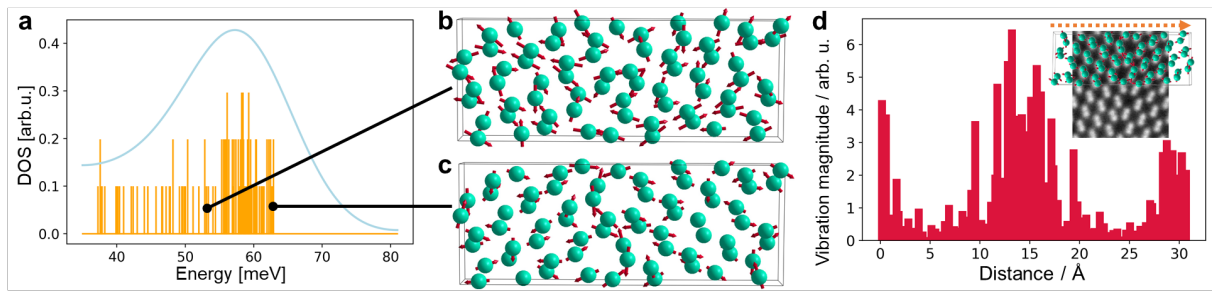

**Fig. 3 | Ab-initio DOS and modes for Σ9 (221). a**, Calculated DOS for the Σ9 (221) GB as histogram plot (orange) and convolved with the experimental energy resolution (blue). **b**, Super cell of the structure in [110] direction (GB running from top to bottom in the center) with arrows representing the atom vibration direction and magnitude (exaggerated) of the bulk mode marked in the histogram in **a**. **c**, Visualization of a localized mode from the histogram in **a**. **d**, Profile of atomic vibration magnitudes across the GB in the super cell shown in **c** and summed perpendicular. The inset shows an overlay of **c** on top of an experimental HAADF image of the Σ9 (221) GB and the arrow indicates the profile direction.

The changes we observe are localized at individual atomic columns as shown in Fig. 2 and Fig. S2 and Fig. S3. The Σ9 (221) boundary has one 5-7-member ring pair per repeat distance while the asymmetric GB has two. Although bond lengths are slightly changed in the 5-7-member rings, there are significant changes in the bond angles. As shown by Rez et al.[21] (Fig. S2 in that article) this leads to a reduced amplitude of the displacements related to the 60 meV optic mode. Given that the bond lengths are minimally changed in the boundary region it is not surprising that there are such small shifts in the peak energy. Furthermore, we see no evidence of peaks that might be due to interface states in the band gap in spectra from the boundary region (Fig. S4). Recent reviews [24,25] show that since all boundary atoms are fourfold coordinated there should be no interface states at the boundary. The Σ3 boundary is coherent, a perfect twin, with the atomic structure unchanged right up to the boundary plane and thus produces virtually no modification of the phonon DOS. However, the same could be said for the stacking fault in 5H SiC investigated by Yan et al. where an enhancement of the acoustic mode intensity was observed when the probe was positioned at the stacking fault [9]. A more careful analysis shows that this fault was in fact a two-layer region of a different polytype, 3C SiC, and that the change in phonon DOS could therefore be explained by the difference in DOS between the two polytypes, so it is likely that these measurements are a consequence of a bulk property. It is interesting to note that the acoustic mode intensity at 40 meV is enhanced for some atoms on either side of the Σ9 (221) boundary plane as shown in Fig. S3b.

The excellent agreement between experiment and simulation gives us high confidence in the predicted modes. Fig. 3a depicts the DOS for the Σ9 (221) GB obtained from ab-initio calculations. In Fig. 3b a typical bulk mode is depicted using arrows indicating the movement of each atom in the super cell of the calculation. A localized mode is shown in Fig. 3c; only the atoms constituting the boundary vibrate strongly, while the other (bulk) atoms exhibit negligible movement. It should be noted that the super cell exhibits a GB not only in the center, but also at the left and right edges. To stress the localization of the mode, Fig. 3d shows a profile of vibration magnitudes perpendicular to the GB and averaged along it with the inset showing the super cell overlaid to the HAADF image and the sketched profile direction. Large vibration amplitudes only exist where the GB atoms are located (center and both edges). As this mode is solely supported by the GB atoms, it should propagate with little damping. Therefore, the GB would act as a waveguide for such a localized mode, allowing it to be used in phononic devices.

We have demonstrated measurements of phonon DOS at atomic resolution at grain boundaries in silicon. Both from experimental STEM-EELS measurements and molecular-dynamics simulations it could be established that changes in the optical phonon modes at GBs are associated with significant changes in bond angles (not lengths) as occurring in 5-fold and 7-fold rings. Calculations of the relevant modes show that this arises from neighboring atom bonds being no longer aligned. We expect that this will apply to similar materials where measurements with high spatial resolution combined with high energy resolution will be more challenging. Our self-consistent scheme to determine and validate local DOS at the atomic scale by means of experiment and simulation should prove a powerful approach for the development of phononics.

Supporting Information

Detailed methods and additional discussions of source-size broadening, energy resolution, robustness of EELS background fitting, sample thickness, extended spectra, surface

oxidation, behavior of 40 meV peak, including Supplementary Figures 1−8 and Supplementary References 1-14

Acknowledgements

T.B. and A.S. are supported in part by the Arizona State University start-up funds. A.S. is also funded by the NSF DMR under Grant No. 1906030 and as part of ULTRA, an Energy Frontier Research Center funded by the U.S. Department of Energy (DOE), Office of Science, Basic Energy Sciences (BES), under Award No. DE-SC0021230. The MD simulations were performed using the Extreme Science and Engineering Discovery Environment (XSEDE), supported by National Science Foundation grant number TG-DMR150006; the HPC resources from Research Computing at Arizona State University; and the National Energy Research Scientific Computing Center, a DOE Office of Science User Facility supported by the Office of Science of the U.S. Department of Energy under Contract No. DE-AC02-05CH11231. The authors thank Ondrej L. Krivanek and Andrea Konečná for critical reading of the manuscript.

Contributions

P.R. conceived the study. B.H. conducted the experiments and B.H and P.R. performed the data analysis. T.M.B. performed structure model pre-/processing and T.M.B., P.R., C.E. and A.S. performed the theoretical calculations. K.M. performed preliminary experiments. C.T.K. built the atomistic GB models from experimental HAADF-STEM images. C.E. provided some pre-investigated atomistic GB models. J.B. contributed the pre-investigated silicon sample. B.H. and P.R. wrote the manuscript. All authors contributed to and approved the final draft of the manuscript.

REFERENCES

(1) Cahill, D. G.; Ford, W. K.; Goodson, K. E.; Mahan, G. D.; Majumdar, A.; Maris, H. J.; Merlin, R.; Phillpot, S. R. Nanoscale thermal transport. *Journal of Applied Physics* **2003**, *93* (2), 793-818.
(2) Cheng, Z.; Li, R.; Yan, X.; Jernigan, G.; Shi, J.; Liao, M. E.; Hines, N. J.; Gadre, C. A.; Idrobo, J. C.; Lee, E.; et al. Experimental observation of localized interfacial phonon modes. *Nat Commun* **2021**, *12* (1), 6901.
(3) Kim, S. I.; Lee, K. H.; Mun, H. A.; Kim, H. S.; Hwang, S. W.; Roh, J. W.; Yang, D. J.; Shin, W. H.; Li, X. S.; Lee, Y. H.; et al. Thermoelectrics. Dense dislocation arrays embedded in grain boundaries for high-performance bulk thermoelectrics. *Science* **2015**, *348* (6230), 109-114.
(4) Zivari, A.; Stockill, R.; Fiaschi, N.; Gröblacher, S. Non-classical mechanical states guided in a phononic waveguide. *Nature Physics* **2022**, *18* (7), 789-793.
(5) Krivanek, O. L.; Lovejoy, T. C.; Dellby, N.; Aoki, T.; Carpenter, R. W.; Rez, P.; Soignard, E.; Zhu, J. T.; Batson, P. E.; Lagos, M. J.; et al. Vibrational spectroscopy in the electron microscope. *Nature* **2014**, *514* (7521), 209-214.
(6) Hage, F.; Kepaptsoglou, D. M.; Ramasse, Q. M.; Allen, L. J. Phonon Spectroscopy at Atomic Resolution. *Phys. Rev. Lett.* **2019**, *122* (1), 016103.

(7) Venkatraman, K.; Levin, B. D. A.; March, K.; Rez, P.; Crozier, P. A. - Vibrational spectroscopy at atomic resolution with electron impact scattering. *Nature Physics* **2019**, *15* (12), 1237-1241.
(8) Hage, F. S.; Radtke, G.; Kepaptsoglou, D. M.; Lazzeri, M.; Ramasse, Q. M. Single-atom vibrational spectroscopy in the scanning transmission electron microscope. *Science* **2020**, *367* (6482), 1124-1127.
(9) Yan, X.; Liu, C.; Gadre, C. A.; Gu, L.; Aoki, T.; Lovejoy, T. C.; Dellby, N.; Krivanek, O. L.; Schlom, D. G.; Wu, R.; et al. Single-defect phonons imaged by electron microscopy. *Nature* **2021**, *589* (7840), 65-69.
(10) Gadre, C. A.; Yan, X.; Song, Q.; Li, J.; Gu, L.; Huyan, H.; Aoki, T.; Lee, S. W.; Chen, G.; Wu, R.; et al. Nanoscale imaging of phonon dynamics by electron microscopy. *Nature* **2022**, *606* (7913), 292-297.
(11) Qi, R.; Shi, R.; Li, Y.; Sun, Y.; Wu, M.; Li, N.; Du, J.; Liu, K.; Chen, C.; Chen, J.; et al. Measuring phonon dispersion at an interface. *Nature* **2021**, *599* (7885), 399-403.
(12) Li, Y. H.; Qi, R. S.; Shi, R. C.; Hu, J. N.; Liu, Z. T.; Sun, Y. W.; Li, M. Q.; Li, N.; Song, C. L.; Wang, L.; et al. Atomic-scale probing of heterointerface phonon bridges in nitride semiconductor. *Proc Natl Acad Sci U S A* **2022**, *119* (8).
(13) Hoglund, E. R.; Bao, D. L.; O'Hara, A.; Pfeifer, T. W.; Hoque, M. S. B.; Makarem, S.; Howe, J. M.; Pantelides, S. T.; Hopkins, P. E.; Hachtel, J. A. Direct Visualization of Localized Vibrations At Complex Grain Boundaries. *Adv Mater* **2023**, e2208920.
(14) Stoffers, A.; Ziebarth, B.; Barthel, J.; Cojocaru-Miredin, O.; Elsasser, C.; Raabe, D. Complex Nanotwin Substructure of an Asymmetric Sigma9 Tilt Grain Boundary in a Silicon Polycrystal. *Phys Rev Lett* **2015**, *115* (23), 235502.
(15) Ziebarth, B.; Mrovec, M.; Elsässer, C.; Gumbsch, P. Interstitial iron impurities at grain boundaries in silicon: A first-principles study. *Physical Review B* **2015**, *91* (3).
(16) Kresse, G.; Furthmuller, J. Efficient iterative schemes for ab initio total-energy calculations using a plane-wave basis set. *Phys. Rev. B* **1996**, *54* (16), 11169-11186.
(17) Togo, A.; Tanaka, I. First principles phonon calculations in materials science. *Scripta Materialia* **2015**, *108*, 1-5.
(18) Frank, W.; Elsässer, C.; Fähnle, M. Ab initio Force-Constant Method for Phonon Dispersions in Alkali Metals. *Physical Review Letters* **1995**, *74* (10), 1791-1794.
(19) Kresse, G.; Furthmüller, J.; Hafner, J. Ab initio Force Constant Approach to Phonon Dispersion Relations of Diamond and Graphite. *Europhysics Letters (EPL)* **1995**, *32* (9), 729-734.
(20) Tersoff, J. Modeling solid-state chemistry: Interatomic potentials for multicomponent systems. *Phys Rev B Condens Matter* **1989**, *39* (8), 5566-5568.
(21) Rez, P.; Boland, T.; Elsässer, C.; Singh, A. Localized Phonon Densities of States at Grain Boundaries in Silicon. *Microsc Microanal* **2022**, *28* (3), 672-679.
(22) Rez, P.; Singh, A. Lattice resolution of vibrational modes in the electron microscope. *Ultramicroscopy* **2021**, *220*, 113162.
(23) Zeiger, P. M.; Rusz, J. Efficient and Versatile Model for Vibrational STEM-EELS. *Phys Rev Lett* **2020**, *124* (2), 025501.
(24) Sun, L.; Marques, M. A. L.; Botti, S. Direct insight into the structure-property relation of interfaces from constrained crystal structure prediction. *Nat Commun* **2021**, *12* (1), 811.
(25) Kohyama, M. Computational studies of grain boundaries in covalent materials. *Modelling and Simulation in Materials Science and Engineering* **2002**, *10* (3), R31-R59.

# Supplementary Information

## for

## Atomic resolution mapping of localized phonon modes at grain boundaries

Benedikt Haas, Tara M. Boland, Christian Elsässer, Arunima K. Singh, Katia March, Juri Barthel, Christoph T. Koch, Peter Rez

Corresponding author: haas@physik.hu-berlin.de

METHODS

Materials and sample preparation

The silicon specimen used in these experiments was the same as that used by Stoffers et al. to investigate the structure of grain boundaries by high-resolution electron microscopy [1]. It was prepared from a Si ingot using an FEI Helios Nanolab 600 dual-beam focused ion beam (FIB). The sample was further thinned and cleaned with an Ar ion beam in a Fischione Nanomill 1040 at 900 eV and 500 eV beam energy, respectively. The thickness over the region shown was measured to be 50 nm from an EELS thickness map (Fig. S5).

Spectrum and image acquisition

Spectra and images were recorded using a Nion HERMES STEM operated at 60 kV, equipped with a Nion IRIS spectrometer fitted with a Dectris ELA direct electron hybrid-pixel detector. High-resolution HAADF images were acquired with a probe defined by a 36 mrad convergence semi-angle, 11 pA beam current and an inner detector angle of 70 mrad. Spectrum images were acquired as 40 maps of 128 by 128 scan points over an 8 nm wide field of view for the Σ3 (111) and symmetric Σ9 (221) GBs and as 30 maps of 128 by 128 points over a 6 nm field of view for the asymmetric (111)|(115) GB. The dwell time was 1 ms per pixel in all cases. The convergence semi-angle was 30 mrad and the current of 220 pA was reduced to 4 pA after monochromation. The spectrometer entrance aperture was 2 mm which corresponded to 42 mrad semi-angle giving an energy resolution of 7.2 meV as defined by the FWHM of the ZLP recorded in vacuum (cf. Fig. S6).

The higher beam current (before the monochromator), means that the source size is larger compared to the high-resolution imaging mode, leading to reduced spatial resolution. This can be determined from a comparison between the HAADF images acquired in parallel with the spectral data and the high-resolution HAADF data. The difference in conditions has the same effect as a broadening (convolution) of the probe with a Gaussian of about 1.0 Å standard deviation (cf. Fig. S1). Spectra were obtained over 1026 pixels with a dispersion of 1 meV per pixel.

Spectrum processing and fitting

For the stacks of spectrum images, the ZLP was aligned within Nion Swift [2] to account for jitter between scan positions but also for slow drift between frames. To correct for spatial drift, the series were subsequently non-rigidly registered and integrated using SmartAlign [3]. The resulting spectrum maps were convolved with a three-dimensional Gaussian function with a standard deviation in both spatial dimensions of 62.5 pm and 2 meV in energy, resulting in smoothed spectra that could be reliably analyzed on a single pixel (single scan position) level. The EELS background was then subtracted for each scan position using a power law fit to two windows in the ranges from 10 meV to 33 meV and from 250 meV to 350 meV using our own Python routines. The robustness of the data and the background subtraction is shown in Fig. S7 below. The spectrum for each scan position was multiplied with the integrated ZLP intensity (between -15 meV and +15 meV) to (partly) compensate for sample thickness variations and the difference in signal for on-column and off-column beam positions. Phonon peak heights and positions were measured by fitting a parabola to the five data points around the peaks (initially determined as maxima in energy ranges) and normalized.

Theoretical modelling

Initial atomic models for the Σ3 (111) were obtained by matching the lateral positions of atomic columns to the high-resolution HAADF data using the QSTEM software [4]. The atomic model for the Σ9 (221) was taken from the atomic coordinates by Ziebarth et al. [5] and asymmetric (111)|(115) boundary was taken from the atomic coordinates used by Stoffers et al. [13] which were checked for agreement with our HAADF data. The initial atomic coordinates were the starting atomic configurations for structural relaxation using molecular dynamics (MD). The phonon DOS were calculated by enumerating the modes obtained from taking the Fourier transform of the velocity–velocity correlation function of the atoms in an extended MD simulation. The MD simulations were performed with the LAMMPS code [6] where the interaction among atoms was described using the Tersoff potential [7] with periodic boundary conditions applied in all directions.

The extended MD simulations were performed using the NPT fix command coupled to a Nosé–Hoover style thermostat and barostat [8-11] to maintain the system at 300 K and 1 atmosphere. A time step of 1 fs was used with a total simulation time of 0.05 ns (50000 steps) where the atom positions were saved at each time step. Additional convergence checks were performed for simulation times of 0.1 ns. These computational settings were sufficient to obtain a stable spectrum. The spectra were calculated using the procedures described by Rez et al.[12]. The trajectories were divided into 24 blocks of 2048 times steps spectra and broadened by a Gaussian with a half width of 9 meV to match the experimental measurements. In calculating the phonon DOS, we made a distinction between those atoms at the GB core and those that were in regions that can be considered to have bond configurations similar to the bulk Si crystal to obtain "GB" and "bulk" phonon DOS.

SUPPLEMENTARY FIGURES

Source-size broadening in spectroscopy

Fig. S1a-c shows a comparison of the high-resolution HAADF images of the three GBs and Fig. S1g-i the HAADF data acquired in parallel during the spectroscopy. A Gaussian blur of the high-resolution data by approximately 1.0 Å is given in Fig. S1d-f and shows good agreement with the spectroscopy images. Thus, the structures are identical and the different appearance originates from the increased source size for the EELS acquisition.

Template-matching for Σ9 (221) GB

To get a clearer picture of the mode suppression and its relation to structure for the symmetric Σ9 (221) we performed template-matching. Fig. S2a depicts the HAADF image from the spectroscopy experiment as shown in Fig. 2b, but with an area marked with an orange outline that was used as a template. A cross-correlation between the template and the full image revealed equivalent areas that were subsequently averaged. In Fig. S2b the map of the 60 meV peak intensity is shown like in main text Fig. S2e (in a more suitable color scale for this purpose). Fig. S2c-e show the result of template matching for the HAADF, the spectroscopy data and a superposition of the two, respectively. It can be seen that atomic scale variations of the mode suppression exist also here and are related to structural features (7-rings).

Behavior of 40 meV peak

Proof that the reduction of the intensity of the 60 meV peak at the GB is not just a geometric effect of the beam mostly going through the "hole" in the atomic structure (7-ring) is given in Fig. S3. In Fig. S3a-c the intensity of the 40 meV peak is depicted for the three GBs, similar to Fig. S2d-f. From the scale it become obvious that the relative change in intensity of this peak is significantly lower than for the 60 meV. In addition, the symmetric Σ9 (221) GB also shows enhancement of the 40meV peak in certain periodic regions, in contrast to the 60 meV peak. In Fig. S3d-f the ratio of 60 meV peak to 40 meV peak is given. This also demonstrates that the suppression of intensity is really related to the local DOS and not to geometric effects.

Extended spectra

Fig. S4 shows averaged raw spectra that show almost no difference between bulk and GB in the region above 100 meV to over 800 meV. Any dangling bonds would have given rise to band gap states in this spectral region. The region below 200 meV was investigated in more detail in the context of surface oxidation below.

Sample thickness

Information about the sample thickness was obtained via mapping of the thickness in terms of the mean free path. Fig. S5a shows a bright-field image of the sample with the positions of the EELS measurements on the three different GBs indicated (cf. main text Fig. 1) and an orange box indicating the area of an EELS thickness map given in Fig. S5b. The map was collected at 60 kV using 10 mrad convergence angle and 42 mrad collection angle. For the area of the Σ3 (111) a value of 0.72 λ (electron mean free path) was obtained. The spectroscopy regions of the symmetric Σ9 (221), and asymmetric (111)|(115) yielded 0.68 λ and 0.57 λ, respectively. Using a value for λ of Si at 200 kV from Egerton [13] and scaling it to 60 kV via the ratio of kinetic energy yields 60 nm. Applying this value, we obtain thicknesses of 43 nm, 41 nm and 34 nm for Σ3, Σ9 and (111)|(115) GB measurements, respectively, with an uncertainty of approximately 10 %.

Energy resolution

Fig. S6 depicts a spectrum obtained in vacuum with the probe parameters as given in the Materials and Methods section. A spline fit yielded 7.2 meV FWHM.

Robustness of background subtraction

Fig. S7 shows the raw data and the robustness of the background subtraction. In Fig. S7a an HAADF image of the symmetric Σ9 (221) is shown with three colored boxes indicating regions of averaged spectra. Fig. S7b depicts the averaged raw spectra from the three regions in comparison and shows that the suppression of the 60 meV peak is already clearly visible in the raw data. Background subtraction by fitting a power law to the tail of the zero-loss peak is shown in Fig. 7c with a single window from 20 meV to 40 meV, instead of the two windows from 10 to 33 meV and 250 to 350 meV used in the article and described in Methods, yielding similar results. This demonstrates the robustness of the background subtraction.

Influence of surface oxidation

To study the effect of surface oxidation, core-loss data was acquired for the (111)|(115) GB. Fig. S8a shows an HAADF of the GB that was acquired together with a spectral map and Fig. S8b depicts the averaged spectrum of the region. It is obvious that next to the strong Si L edge, only very small amounts of C from surface contamination and tiny amounts of oxygen are present. Fig. S8c-d show the distribution of Si and O as elemental maps. The darker regions in the Si (cf. color scale bar) originate from sputtering of the Si at the more weakly bonded defect atoms. The acquisition parameters were 60 kV primary energy and 262 s exposure to 250 pA, thus a much higher dose than for the vibrational spectra. Fig. 8e shows the HAADF of the (111)|(115) GB from the main text with three boxes indicating regions of spectrum averaging. Extended spectra in the region where silicon oxide vibrations exist are shown in Fig. 8f and can be compared to the work of K. Venkatraman et al.[14] investigating Si and $SiO_2$. This demonstrates that the influence of surface oxidation on the data is negligible.

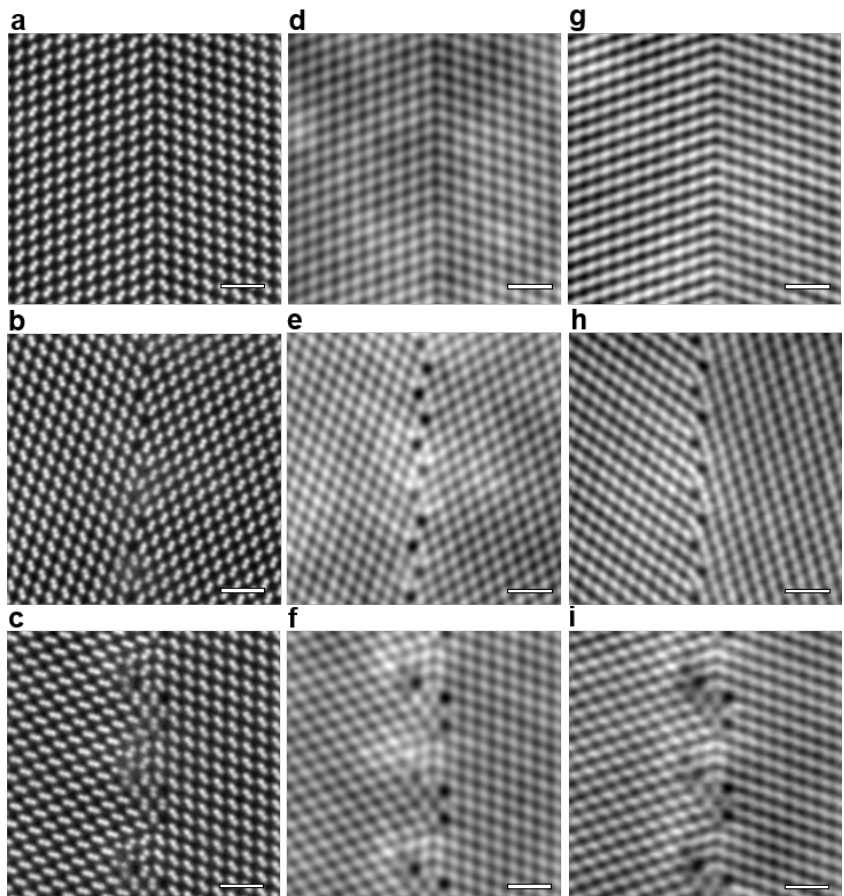

**Fig. S1 | Comparison of high-resolution HAADF and HAADF from spectral maps. a-c**, High-resolution HAADF images of Σ3 (111), symmetric Σ9 (221), and asymmetric (111)|(115) GBs, respectively. Scale bars, 1

nm. **d-f**, Convolution of the high-resolution HAADF images with a Gaussian of about 1.0 Å. Scale bars, 1 nm. **g-i**, HAADF images acquired together with the spectrum maps. Scale bars, 1 nm.

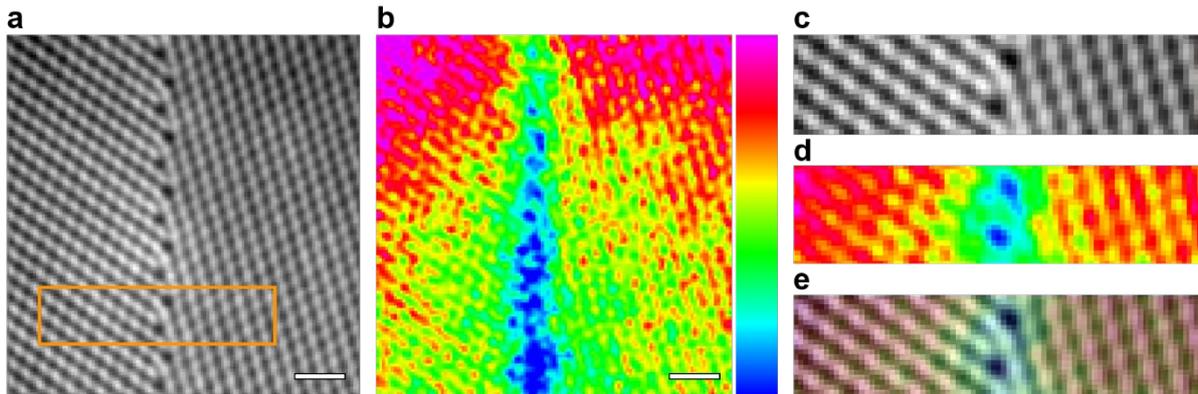

**Fig. S2 | Template-matching for the symmetric Σ9 (221) GB. a**, HAADF image acquired together with the spectroscopy data as in Fig. 2b but with an orange rectangle marking the region used as template for template-matching. Scale bar, 1 nm. **b**, Map of the 60 meV peak intensity as in Fig. 2e but with a different color scale. Scale bar, 1 nm. **c**, Result from averaging similar regions of the HAADF image via template-matching. **d**, Result of the template-matching applied to the map in **b**. **e**, Overlay of semi-transparent spectroscopy data **d** onto HAADF data **c**.

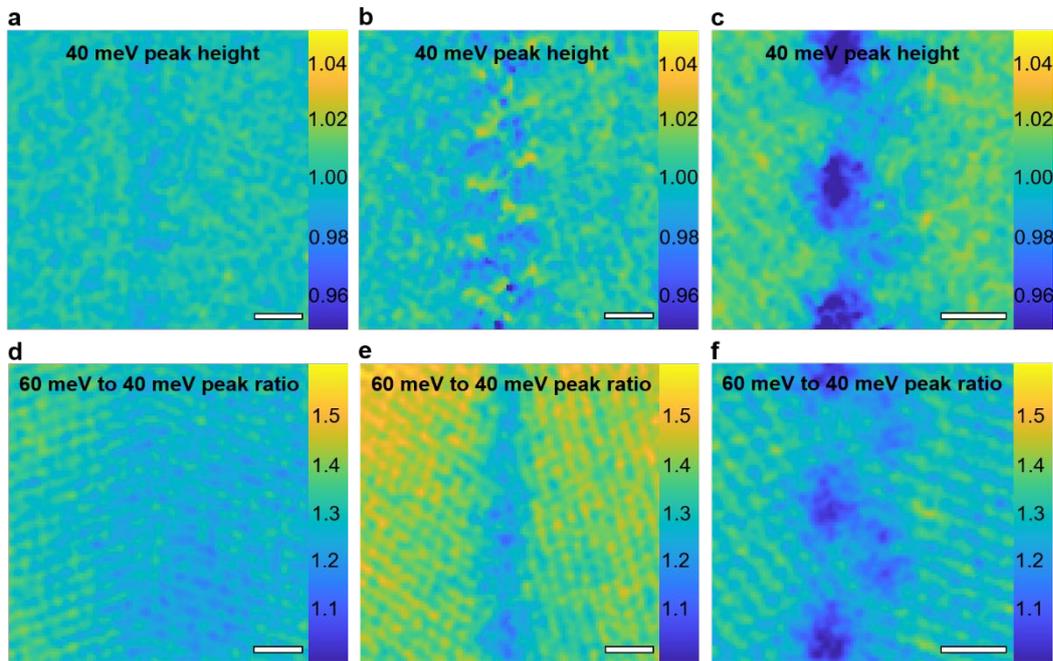

**Fig. S3 | Maps of the 40 meV peak and ratio of 60 meV to 40 meV. a-c**, Maps for the intensity of the 40 meV peak for Σ3 (111), symmetric Σ9 (221), and asymmetric (111)|(115) GBs, respectively. Scale bars, 1 nm. **d-f**, Maps of the ratio of 60 meV peak to 40 meV peak for the three GBs. Note the small enhancement of the 40 meV acoustic mode at points on either side of the Σ9 (221) GB in **b**. Scale bars, 1 nm.

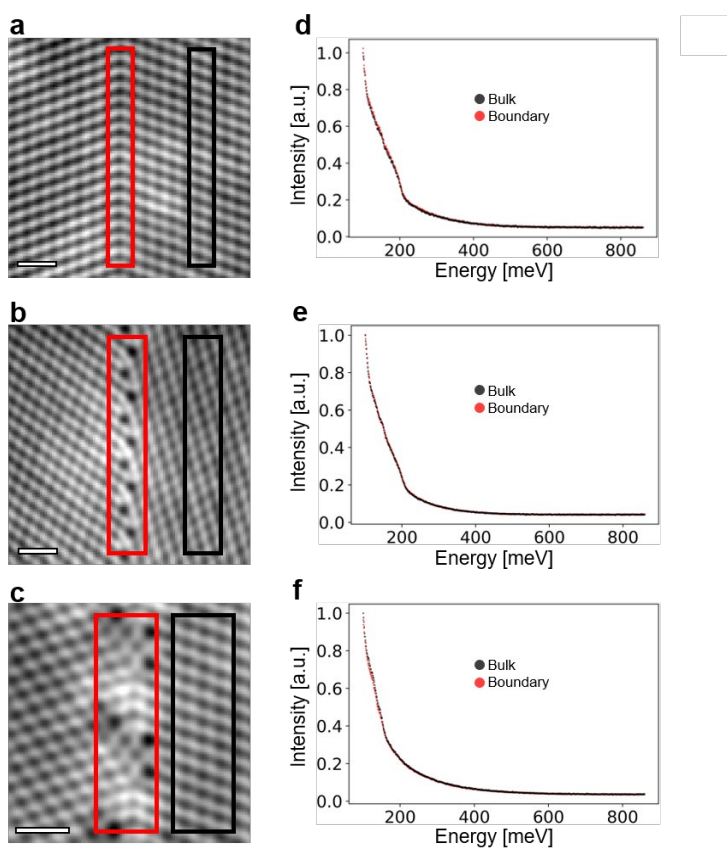

**Fig. 4 | Extended spectra. a-c**, HAADF data for the Σ3 (111), symmetric Σ9 (221), and asymmetric (111)|(115) GBs, respectively, with black and red boxes indicating the regions of GB and bulk used for the spectra in **d-f**. Scale bars, 1 nm. **d-f**, Averaged raw spectra of the indicated regions between 100 meV and 860 meV, showing no major differences between bulk and GB regions.

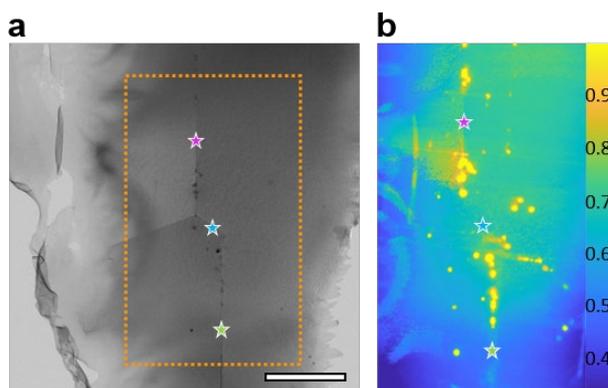

**Fig. S5 | Thickness map of the sample. a**, Bright-field STEM image of the utilized sample with the positions of the EELS measurements of the different GBs indicated as colored stars. Scale bar, 500 nm. **b,** Thickness map in units of electron mean free path λ acquired over the region indicated by the orange dashed line in **a**. Carbon contamination from previous measurements lead to spots of high values.

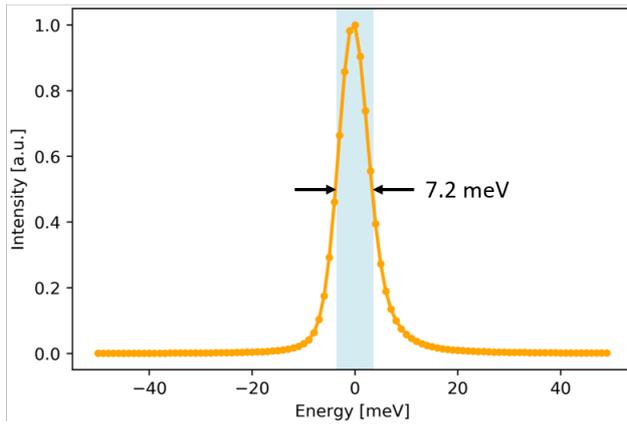

**Fig. S6 | Energy resolution.** Zero-loss peak of the 30 mrad probe in vacuum (cf. Methods section for details) with the FWHM of 7.2 meV (determined by spline fitting) as a measure for the energy resolution indicated.

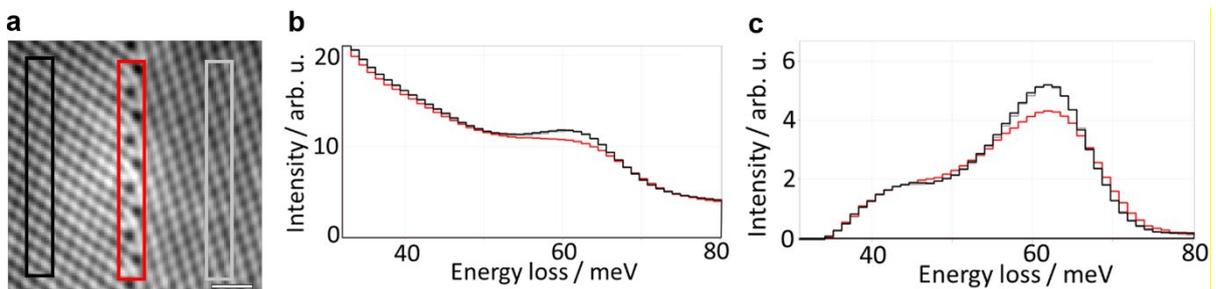

**Fig. S7 | Robustness of background subtraction. a**, HAADF image of symmetric Σ9 (221) GB (as shown in the main text) with three colored boxes indicating areas of averaged spectra shown in **b** and **c**. Scale bar, 1 nm. **b**, raw spectra showing already clearly the suppression of the 60 meV peak at the boundary in comparison to the left and right bulk. **c**, spectra after subtracting the ZLP tail via power law fit with a single window from 20 to 40 meV.

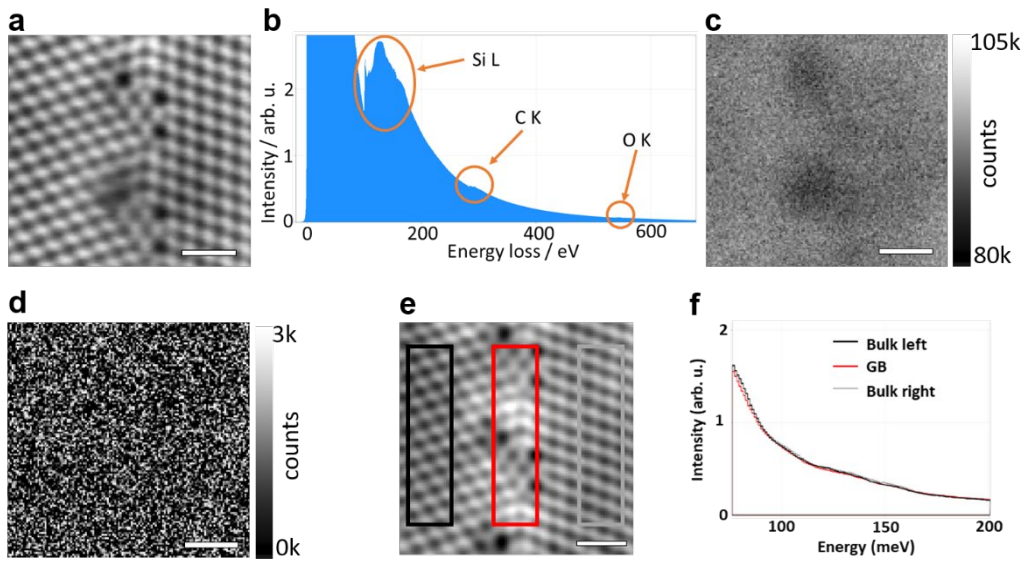

**Fig. S8 | Investigation of surface oxidation. a**, HAADF data for the asymmetric (111)|(115) GB acquired together with a core-loss spectral map. Scale bar, 1 nm. **b**, Averaged spectrum over the depicted area showing the strong Si L edge together with a small C K edge (surface contamination) and a tiny O K (surface oxidation). **c**, map of the Si L electrons. Scale bar, 1 nm. **d**, map of the O K edge. Scale bar, 1 nm. **e**, HAADF of the asymmetric (111)|(115) GB (as shown in the main text) with three colored boxes indicating averaged regions. Scale bar, 1 nm. **f**, Averaged spectra from the regions indicated in **e**.

Supplementary References

(1) Stoffers, A.; Ziebarth, B.; Barthel, J.; Cojocaru-Miredin, O.; Elsasser, C.; Raabe, D. Complex Nanotwin Substructure of an Asymmetric Sigma9 Tilt Grain Boundary in a Silicon Polycrystal. *Phys Rev Lett* **2015**, *115* (23), 235502.
(2) Meyer, C.; Dellby, N.; Hachtel, J. A.; Lovejoy, T.; Mittelberger, A.; Krivanek, O. Nion Swift: Open Source Image Processing Software for Instrument Control, Data Acquisition, Organization, Visualization, and Analysis Using Python. *Microscopy and Microanalysis* **2019**, *25* (S2), 122-123.
(3) Jones, L.; Varambhia, A.; Beanland, R.; Kepaptsoglou, D.; Griffiths, I.; Ishizuka, A.; Azough, F.; Freer, R.; Ishizuka, K.; Cherns, D.; et al. Managing dose-, damage- and data-rates in multi-frame spectrum-imaging. *Microscopy (Oxf)* **2018**, *67* (suppl_1), i98-i113.
(4) Koch, C. T. Determination of core structure periodicity and point defect density along dislocations. Arizona State University, 2002.
(5) Ziebarth, B.; Mrovec, M.; Elsässer, C.; Gumbsch, P. Interstitial iron impurities at grain boundaries in silicon: A first-principles study. *Physical Review B* **2015**, *91* (3).
(6) Plimpton, S. Fast Parallel Algorithms for Short-Range Molecular Dynamics. *Journal of Computational Physics* **1995**, *117* (1), 1-19.
(7) Tersoff, J. Modeling solid-state chemistry: Interatomic potentials for multicomponent systems. *Phys Rev B Condens Matter* **1989**, *39* (8), 5566-5568.
(8) Nosé, S. A unified formulation of the constant temperature molecular dynamics methods. *The Journal of Chemical Physics* **1984**, *81* (1), 511-519.
(9) Hoover, W. G. Canonical dynamics: Equilibrium phase-space distributions. *Phys Rev A Gen Phys* **1985**, *31* (3), 1695-1697.
(10) Martyna, G. J.; Tobias, D. J.; Klein, M. L. Constant pressure molecular dynamics algorithms. *The Journal of Chemical Physics* **1994**, *101* (5), 4177-4189.
(11) Müser, M. H.; Binder, K. Molecular dynamics study of the α-β transition in quartz: elastic properties, finite size effects, and hysteresis in the local structure. *Physics and Chemistry of Minerals* **2001**, *28* (10), 746-755.
(12) Rez, P.; Boland, T.; Elsässer, C.; Singh, A. Localized Phonon Densities of States at Grain Boundaries in Silicon. *Microsc Microanal* **2022**, *28* (3), 672-679.
(13) Egerton, R. F. *Electron Energy Loss Spectroscopy in the Electron Microscope*; Plenum, 1986.
(14) Venkatraman, K.; Rez, P.; March, K.; Crozier, P. A. The influence of surfaces and interfaces on high spatial resolution vibrational EELS from SiO2. *Microscopy (Oxf)* **2018**, *67* (suppl_1), i14-i23.